\documentclass[journal]{IEEEtran}
\usepackage[dvips]{color} 
\usepackage[dvips]{graphicx}
\ifCLASSINFOpdf
\usepackage[pdftex]{graphicx}
\else
\fi
\begin{document}
\title{Magnetic structure of domain walls\\ confined in a nano-oxide layer}
\author{Katsuyoshi~Matsushita\thanks{Katsuyoshi Matsushita, Email: k-matsushita@aist.go.jp}, 
Jun~Sato and
Hiroshi~Imamura\\
Nanotechnology Research Institute (NRI),\\
Advanced Industrial Science and Technology (AIST),\\
AIST Tsukuba Central 2, Tsukuba, Ibaraki 305-8568, Japan.}


\markboth{Journal of \LaTeX\ Class Files,~Vol.~6, No.~1, January~2007}%
{Shell \MakeLowercase{\textit{et al.}}: Bare Demo of IEEEtran.cls for Journals}



\maketitle

\begin{abstract}
%\boldmath
In the recent years, a spin-valve was developed with a current-confined-path structure consisting of a non-oxide-layer (NOL).
We analyze magnetic structures of the current-confined-path 
in the nano-oxide layer sandwiched between ferromagnetic electrodes
and clarify the dependency of the magnetic structure on the shape
and size of the current-confined-path. 
Our results of stiffness energy density and thermal stability shows
that we should fabricate a CPP spin-valve with narrow current-confined-paths with large
aspect ratio and strong constriction in order to enhance the MR ratio. 
\end{abstract}


% \fi
\IEEEpeerreviewmaketitle

%\section{Introduction}
\IEEEPARstart{G}{eometrically} confined domain walls of nanometer size
contacts have attracted enormous attention of researchers because of
its possibility in applications to future spintronic
devices~\cite{Chopra}-\cite{Levy}.  In the last decade much effort
has been devoted to study magnetoresistance (MR) of domain walls in atomic wires.  
One of main obstacles to applications of the
atomic wires is mechanical stability.  For example, Chopra {\it et
al}.~\cite{Chopra} reported that a Co atomic wire that shows 300\%
MR ratio is stable for periods of up to as short
as 2-3 mins.

Recently, Fuke {\it et al}. fabricated a spin-valve of magnetic
nanomultilayers with a current-confined-path structure consisting of a
nano-oxide-layer (NOL) with a lot of fine holes filled with
ferromagnetic metal~\cite{Fuke,Doi}.  The holes of an NOL are very
stable and its typical size is as small as a few nanometers.  They
showed that MR ratio of the spin-valve increases with increasing a
value of resistance area product, which means that the MR ratio due to
the geometrically confined domain walls is enhanced for the narrow
metallic channels.  In order to understand these experimental results,
it is important to study magnetic structures of the domain walls
confined in the NOL.  However, it is very difficult to observe the
magnetic structure of the domain walls directly by using a scanning
probe microscope or optical techniques because the domain
walls are confined in the NOL.

The micromagnetic simulation technique provides a powerful tool for
studying magnetic properties of such geometrically confined domain
walls~\cite{Coey}-\cite{Komine2}.
Magnetic structures of the confined domain walls are investigated in atomic wire bridges with crystalline anisotropy by several groups~\cite{Coey}-\cite{Molyneux}.
However the effect of the crystalline anisotropy on magnetic structures should be small for the NOL spin-valve~\cite{Fuke,Doi} where the size of the domain walls is of the order of a few nanometers.
Although a few simulations for contacts without crystalline anisotropy have been reported in \cite{Gorkom}-\cite{Komine2},
a certain careful simulation where the real shape of the current-confine-path in the NOL spin-valve\cite{Fuke,Doi} is taken into account is needed to design the highly sensitive magnetoresistive devices.

In the present paper, we study shape dependence of magnetic
structures of current-confined-paths in the NOL by using the micromagnetic
simulation. 
We evaluate stiffness energy density of the magnetic structures in the
NOL sandwiched between two magnetic layers of which magnetizations are
aligned antiparallel to each other.
We can estimate the effective thickness of the geometrically confined
domain wall from the calculated stiffness energy density.
We show that a narrow current-confined-path with large 
aspect ratio and strong constriction is required in order to enhance the MR ratio. 

We employ a spin model on the simple cubic lattice, which is defined by the Hamiltonian,
\begin{eqnarray}
&{\cal H} &= - J \sum_{\left<i,j\right>} \vec S_{i} \cdot \vec S_{j} \nonumber \\ 
&&+ K_d\sum_{i} \vec S_{i} \cdot \int d\vec r \hat {\cal D}(\vec x_{i} - \vec r) \cdot \vec S(\vec r) \nonumber \\ 
&&+ K_{u}\sum_{i}(1-(\vec e_i \cdot \vec S_{i})^2) + \sum_{i} \vec h_{i} \cdot \vec S_{i}. \label{Hamiltonian}
\end{eqnarray}
The local values at the $i$-th site, $\vec S_{i}$, $\vec x_i$, $\vec
e_i$ and $\vec h_i$ denote an unit vector representing a classical
Heisenberg spin, a coordinate, an unit vector aligned in a direction
of crystalline anisotropy and an applied magnetic field, respectively. 
The first term in the right hand side of (\ref{Hamiltonian}) is
the exchange stiffness energy between spins at nearest neighbor
sites. The exchange coupling constant is denoted by $J$. The value of $J$ is
related to exchange stiffness constant, $A$, and a lattice constant,
$a$, as $J = 2aA$.

The second term is the dipole-dipole interaction energy.  The tensor
function, $\hat {\cal D}(\vec x)$, is expressed as,
\begin{eqnarray}
\hat {\cal D}(\vec x) = \frac{1}{4\pi} \frac{1-3\vec e_{\vec x} \otimes \vec e_{\vec x} }{|\vec x|^3}, \nonumber
\end{eqnarray}
where $\vec e_{\vec x} = \vec x/|\vec x|$.
The dipole-dipole coupling constant, $K_d$, is equivalent to
$a^3\mu_0M_s^2/2$, where $M_s$ and $\mu_0$ are the saturation
magnetization and vacuum permeability, respectively.
In order to deal with the dipole-dipole interaction in the NOL spin-valve, 
we adopt a finite element - boundary element (FEM-BEM) hybrid
method~\cite{Fredkin}.
In the evaluation of the stray field coming from the dipole-dipole interaction,
spins at lattice points are extrapolated to that in the whole region in the NOL, $\vec S(\vec r)$. 

The third term is the crystalline anisotropic energy.
We neglect this term in the present calculation
because the characteristic length determined by the crystalline anisotropy 
is a few-tens of nanometers and therefore the effect of the crystalline
anisotropy on the magnetic structure of the geometrically confined
domain wall of a few nanometer size must be very small.

The forth term is the Zeeman energy due to the applied magnetic field. 
We also neglect the Zeeman energy in the present calculation and the
direction of the magnetizations in the top and bottom electrodes are
determined by the boundary conditions.
The boundary conditions we use are the following:
 $S_i^x= -1$ on top and side surfaces of the top electrode and 
$S_i^x= 1$ on bottom and side surfaces of the bottom electrode. 
On the other surface the boundary condition is free.

The system we consider is schematically shown in Fig.~\ref{fig:Embedded_FEM_model}.
\begin{figure}[tb]
\begin{center}
\end{center}
\caption{
    The finite element mesh of the current-confined-path in the
    NOL spin-valve is shown. 
    The shape of the current-confined-path is deformed to a hyperboloid
    parametrized by two ratios, $d/h$ and $d_{\rm c}/d$, where $d$,
    $d_{\rm c}$ and $h$, denote the diameter of the center, that of the
    bottom and height, of the current-confined-path, respectively. 
}
\label{fig:Embedded_FEM_model}
\end{figure}
The system is divided to a 3-dimensional
lattice consisting of hexahedral finite elements.
The number of elements in the top electrode is taken to be
$39 \times 39 \times 3=4563$.  
The bottom electrode is divided in the same way as in the top electrode.
The number of elements in the $z$-direction in the contact is taken to
be 9. The total number of the elements, which depends on the ratio
$d/h$, is of the order of 10$^4$.  
The shape of the contact is assumed to be a hyperboloid
parametrized by two ratios, where $d$ and $d_{\rm c}$ are the diameters, respectively, at the center and the bottom of the contact, and $h$ is the height of the contact.  
%========================================
% Exchange length
%========================================
The characteristic length determined by the competition between 
the exchange interaction ${\cal H}_{\rm ex}$ and the dipole-dipole
interaction ${\cal H}_{\rm d}$ is the exchange length defined by,
\begin{equation}
l_{\rm ex} = \sqrt{\frac{A}{K_{\rm d}}},
\end{equation}
which is of the order of a few nanometers for conventional
ferromagnetic metals~\cite{Hubert}. We assume that $l_{\rm ex}=$2.8nm corresponding to the materials used in~\cite{Fuke}.
The unit length of the system, $a$, is taken to be $\sqrt{K_d/J}l_{\rm
  ex}$ and the height of the NOL is $h=2$nm ($\sim 0.71l{\rm ex}$). 

The classical Heisenberg spin system is embedded in the whole region as shown in Fig.~\ref{fig:Embedded_FEM_model}.
The spins obeys the Landau-Lifshitz-Gilbert equation,
\begin{eqnarray}
\frac{d}{dt} \vec S_i = \frac{\gamma}{1+{\alpha}^2} \vec S_i \times \{  \frac{\partial {\cal H}}{\partial \vec S_i} + \alpha \vec  S_i \times \frac{\partial {\cal H}}{\partial \vec S_i}\},
\end{eqnarray}
where $\gamma$ and $\alpha$ are the gyromagnetic ratio and the Gilbert damping constant, respectively. 
The calculated ground states by the present simulation do not depend on their values.
We simulates relaxation of the spins from various initial states by
numerically solving the equation at zero temperature to obtain the
ground state.
% The initial spin configurations excp the boundary surface where
% spins are fixed for all of the present simulation are the set of a
% N{\'e}el-wall like state,
% \begin{eqnarray}
% \vec S_i = \left(\cos(2\pi z_i/aL_z), 0 , \sin(2\pi z_i/aL_z) \right),\label{Neel}
% \end{eqnarray}
% and a Bloch-wall like state,
% \begin{eqnarray}
% \vec S_i = \left(\cos(2\pi z_i/aL_{\rm z}), \sin(2\pi z_i/aL_{\rm z}) , 0 \right), \label{Bloch}
% \end{eqnarray}
% Here, $z_i$ are components of the coordinate of the $i$-th site.
The time integration is continued until a space- and
time-averaged value of the torque exerted to the spins is of the order of
10$^{-8}\gamma J/(1+\alpha^2)$.
In the present simulation, further time-integration does not produce
additional change of magnetic structures within the reachable time
scale by our computers.
The obtained lowest energy state by simulations started from three
initial states is regarded as the ground state at each given set of
parameters.
% The choice of the initial states is determined for easily obtaining the ground states mentioned below.

\begin{figure}[tb]
\begin{center}
\end{center}
\caption{
  {
    The magnetization configurations of
    the N{\'e}el-wall like ground state for  $d/h$ = 0.8 is shown in the
    panel (a).
    The magnetization configurations of
    the Bloch-wall like ground state for  $d/h$ = 0.8 is shown in the
    panel (a).
    The diagram of magnetic structures of the ground states is shown in
    the panel (c). The N{\'e}el-wall like , Bloch-wall like structures are indicated by the diamonds, cubes and triangles.
  }
}
\label{fig:magnetic_Structure_diagram}
\end{figure}
In the present simulation, by changing two ratios, $d/h$
and $d_c/d$, we investigate the systems in the range of $0.8 < d/h < 4.0$ and for $d_{\rm c}/d$ = 0.25, 0.5, 0.75 and 1.0. The obtained
 ground states are the N{\'e}el- and Bloch-wall like states as shown in
Figs.~\ref{fig:magnetic_Structure_diagram}(a) and \ref{fig:magnetic_Structure_diagram}(b), respectively.

The magnetic structure diagram is shown in
Fig.~\ref{fig:magnetic_Structure_diagram}(c).  At small values of
$d/h$, the ground states are the N{\'e}el-wall like state.  On the
other hand the ground states observed at large values of $d/h$ are the
Bloch-wall like state.  The similar results were obtained for
atomic wires with a crystalline anisotropy by Coey et al.~\cite{Coey}.  
The result can be understood as follows:
Our results show that the magnetic structures depend almost only on $z$-component of coordinates.
Thus the exchange energy does not depend on whether the Bloch-wall or N\'{e}el-wall like structures \cite{Bruno} and
we can discuss the energy of the states only by the dipole-dipole interaction energy.
The dipole-dipole interaction energy on the surface of the contact, 
which decreases with increasing the aspect ratio, $d/h$, gives a dominant
contribution to the total energy of the Bloch-wall like state.
On the other hand, the dipole-dipole interaction energy in the
interior of the contact, which increases with increasing the aspect ratio, $d/h$, 
gives a dominant contribution to the total energy of the N{\'e}el-wall like state.
Therefore the Bloch-wall (N{\'e}el-wall) like state is preferred for the wide
(narrow) contact as show in Fig.2 (c).
As we increase the ratio $d_c/d$, the dipole-dipole interaction on the surface of the contact increases 
but that in the interior of the contact decreases.  
Therefore the value of the aspect ratio $d/h$ which
determines the boundary between the Bloch- and N{\'e}el-like wall states
is a decreasing function of the ratio $d_c/d$.

\begin{figure}[tb]
\begin{center}
\end{center}
\caption{Concentration of stiffness energy at $d_{\rm c}/d$= 0.5
  $d/h$= 0.8 (a),  $d_{\rm c}/d$= 0.5 $d/h$= 3.9 (b), $d_{\rm c}/d$=
  1.0 $d/h$= 0.8 (c) and  $d_{\rm c}/d$= 1.0 $d/h$= 3.9 (d).}
\label{fig:Stiffness_energy}
\end{figure}

Figs.~\ref{fig:Stiffness_energy}(a)-\ref{fig:Stiffness_energy}(d) show the dependence of the stiffness energy
density on the size and shape of the contact.
The stiffness energy density is defined as
\begin{eqnarray}
E^i_{\rm ex} \equiv \frac{J}{2v_i} \int_{v_i} d\vec r \left( \nabla \vec S(\vec r) \right)^2, \\ 
\sim \frac{J}{2a^2} \int_{z_i+a/2}^{z_i-a/2} dz \left|\nabla \theta(z)\right|^2,\label{eq:Stiffness_energy_estimation}
\end{eqnarray}
where $\theta(z)$ denotes azimuthal angle for the Bloch-wall
like state and polar angle for the N{\'e}el-wall like state.
$v_i$ is total volume of elements near the $i$-th site.
The last expression in (\ref{eq:Stiffness_energy_estimation}) is
justified for much small values of  $K_d/J$. 
In our simulation, value of $K_d/J$ about 0.006 and thereby the approximation is also justified. 

The stiffness energy densities of the narrow contacts are shown
in Figs.~\ref{fig:Stiffness_energy}(a) and \ref{fig:Stiffness_energy}(b). The stiffness energy
densities are almost confined in the NOL and its spacial distribution can
be controlled by changing the shape of the contact.
Comparing the results shown in Figs.~\ref{fig:Stiffness_energy}(a) and
\ref{fig:Stiffness_energy}(b), one can see that the stiffness energy density of $d_{\rm c}/d =
0.5$ is much more concentrated at the center of the system than that of
$d_{\rm c}/d=1.0$. 
Since MR ratio increases with decreasing the thickness of the
domain wall\cite{Levy},  we can expect a larger value of the MR ratio
of the contact with $d_{\rm c}/d$ = 0.5 than that with $d_{\rm c}/d$ = 1.0.

The stiffness energy densities of the wide contacts are shown
in Figs.~\ref{fig:Stiffness_energy}(c) and \ref{fig:Stiffness_energy}(d). 
The stiffness energy densities stick out from the NOL and therefore the
effective thickness of the domain wall is thicker than that for the
system with the narrow contacts.
As shown in Fig.~\ref{fig:Stiffness_energy}(c), a strong
concentration of the stiffness energy density appears around the
surface of the contact with $d_{c}/d$=0.5.
Although  a strong concentration of the stiffness energy density means
that we have a thin domain wall around the surface, we cannot expect
the large MR ratio because the other region with small
stiffness energy density gives a small contribution to the MR ratio and
acts as a parasitic resistance.  

Finally, we discuss the stability of the ground states.  
The bistability of the Bloch and N\'{e}el-like structures is observed for almost all of the region of
the parameters except for $d_c/d$ =0.25 and $d/h \le 3$.
Thus thermal instability is expected in those as discussed in \cite{Coey,Labaye}.  
In our results, the energy difference between the Bloch- and N{\'e}el-wall like states is of the order of a few Kelvins at most. 
Although in wide contacts
the energy difference between the Bloch- and N{\'e}el-wall like states increases with increasing system volume.
% with the
%Bloch-wall like state, energy gain from the excited N{\'e}el-wall like
%state is expected to be proportional to the system volume because the
%main contribution of the difference is the dipole-dipole interaction
%on interior of the system. 
If $h$ is set at the typical experimental value of 2 nm \cite{Fuke},
$d$ should be a few tens nanometers at least in order that the energy difference reaches room temperatures. 
However, as discussed before, we can not expect large MR ratios for such wide contacts.
In the Neel-wall like states, we do not observe the excited Bloch state for $d_c/d$
=0.25 and $d/h \le 3.3$ within our calculation
accuracy.
Therefore the narrow contacts with large aspect ratio and strong constriction has stable magnetization.
The narrow contacts are a potential candidate of highly sensitive magnetoresistive devices because of their high magnetoresistance ratio as mentioned before.
% {\color{red}Therefore, the narrow contacts with large aspect ratio and hyperboloid-like constriction has not only high magnetoresistance as mentioned previously and but also much stable magnetic structure.}

In summary, we analyze magnetic structures of a current-confined-path 
in a nano-oxide layer sandwiched between ferromagnetic electrodes
and clarify the dependency of the magnetic structure on the shape
and size of the current-confined-path. 
Our results of stiffness energy density and thermal stability shows
that we should fabricate narrow current-confined-path with large
aspect ratio and strong constriction in order to enhance the MR ratio. 

The authors thank M.~Sahashi, M.~Doi, H.~Iwasaki, M.~Takagishi,
Y.~Rikitake and K.~Seki for valuable discussions.  The work
has been supported by The New Energy and Industrial Technology
Development Organization (NEDO).  K. M. is partially supported by
Grant-in-Aid for Young Scientist from the Ministry of Education,
Science, Sports and Culture of Japan.

\ifCLASSOPTIONcaptionsoff
  \newpage
\fi


\begin{thebibliography}{1}

\bibitem{Chopra} H.~D. Chopra, M.~R.~Sullivan, J.~N.~Armstrong and S.~Z.~Hua, ``The quantum spin-valve in cobalt atomic point contacts'', Nature~materials vol.~4, pp.~832-837, Oct.~2005. 
\bibitem{Garcia} N.~Garcia, M.~Munoz, and Y.-W.~Zhao, ``Magnetoresistance in excess of 200\% in Ballistic Ni Nanocontacts at Room Temperature and 100 Oe'', Phys.~Rev.~Lett. vol.~82, pp.~2923-2926 Apr.~1999. 
\bibitem{Bruno} P.~Bruno, ``Geometrically Constrained Magnetic Wall'', Phys.~Rev.~Lett., vol.~83, pp.~2425-2428, Sept.~1999.
\bibitem{Imamura} H.~Imamura, N.~Kobayashi, S.~Takahashi, and S.~Maekawa, ``Conductance Quantization and Magnetoresistance in Magnetic Point Contacts'', Phys. Rev. Lett., vol.~84, pp.~1003-1006, Jan.~2000.
\bibitem{Ohsawa} Y. Ohsawa, ``Magnetoresistance Characterization of NiFe Films With a Planar Point Contact'',
IEEE~Trans.~Magn., vol.~43, pp.~3007-3009, Jun.~2007.
\bibitem{Coey} J.~M.~D.~Coey, L.~Berger, and Y.~Labaye, ``Magnetic excitations in a nanocontact'', Phys.~Rev.~B, vol.64, pp. 020407~1-3, Jul.~2001. 
\bibitem{Labaye} Y.~Labaye, L.~Berger, and J.~M.~D.~Coey, ``Domain walls in ferromagnetic nanoconstriction'', J.~Appl.~Phys., vol.~91 pp.~5341-5346, Apr.~2002.
\bibitem{Savchenko} L.~L.~Savchenko, A.~K.~Zvezdin, A.~F.~Popkov and K.~A.~Zvezdin, ``Magnetic configurations in the region of a nanocontact between ferromagnetic bars'', Phys.~Sol.~Stat., vol.~43, pp.~1509-1514, Aug.~2001.
\bibitem{Zvezdin} K.~A.~Zvezdin, A.~V.~Khval'kovskii, ``Phase transformations of the magnetic structure in film nanobridges'',
Phys.~Sol.~Stat.,  vol.~47   pp.~1176-1185, Jun.~2005. 
\bibitem{Molyneux} V.~A.~Molyneux, V.~V.~Osipov, and E.~V.~Ponizovskaya, ``Stable two- and three-dimensional geometrically constrained magnetic structures: The action of magnetic fields'', Phys.~Rev.~B, vol.~65 pp.~184425~1-6, May~2002.
\bibitem{Gorkom} R.~P.~van~Gorkom, A.~Brataas, and G.~E.~W.~Bauer, ``Micromagnetics and magnetoresistance of a Permalloy point contact'', Appl.~Phys.~Lett., vol.~74 pp.422-424, Jan.~1999.
\bibitem{Komine} T.~Komine, T.~Takahashi, R.~Sugita, T.~Muranoi, and Y.~Hasegawa, ``Micromagnetic calculation of the magnetization process in nanocontacts'', J.~Appl.~Phys., vol.~97, pp.~10C508~1-3, May.~2005.
\bibitem{Komine2} T.~Komine, T.~Takahashi, S.~Ishii, R.~Sugita, T.~Muranoi, and Y.~Hasegawa, ``Micromagnetic analysis of a magnetic domain wall in 2-D and 3-D nanocontacts'', IEEE~Trans.~Magn., vol.~41 pp.~2586-2588 Oct.~2005. 
\bibitem{Fuke} H.~N.~Fuke, S.~Hashimoto, M.~Takagishi, H.~Iwasaki, S.~Kawasaki, K.~Miyake and M.~Sahashi,
, ``Magnetoresistance of FeCo Nanocontacts With Current-Perpendicular-to-Plane Spin-Valve Structure '', IEEE~Trans.~Magn. vol.~43 pp.~2848-2850, Jun.~2007.
\bibitem{Doi}  M.~Doi, H.~Endo, K.~Shirafuji, M.~Takagishi, H.~N.~Fuke, H.~Iwasaki, M.~Sahashi, ``Micromwave oscillation study on self-assembling nano-confined domain wall structure'', 52nd MMM conference, CE-15, Nov.~2007.
\bibitem{Levy} P.~M.~Levy and S.~Zhang, Phys.~Rev.~Lett., vol.~79, pp.~5110-5113, Dec.~1997.
\bibitem{Fredkin} D. R. Fredkin and T. R. Koehler, ``Hybrid method for computing demagnetizing fields'',
 IEEE~Trans.~Magn., vol.~26 pp.~415-417, Mar.~1990. 
\bibitem{Hubert} A.~Hubert and R.~Sch\"afer, ``Magnetic Domains'', Springer-Verlag Berlin Heidelberg New York 2000. 
\end{thebibliography}
\end{document}